\documentclass[preprint,cite,showkeys,preprintnumbers,amsmath,amssymb]{revtex4}

\usepackage[dvips]{graphics}
\usepackage{color}
\usepackage{epsfig}

\newcommand{\met}{\rlap{\,/}E_T}

\begin{document}

\title{Phenomenology of Top Partners at the ILC}%


\author{Kyoungchul Kong\footnote{email: kckong@fnal.gov} and Seong Chan Park
\footnote{email: spark@mail.lns.cornell.edu and spark@phya.snu.ac.kr}}
\affiliation{${}^*$
Theoretical Physics Department, Fermilab, Batavia, IL 60510, USA \\
${}^\dag$ FPRD, School of Physics and Astronomy, Seoul National
University, Seoul 151-742, Korea}

\begin{abstract}
The assumption of a new symmetry provides a nice
explanation of the existence of dark matter and an elegant way to
avoid the electroweak constraints. This symmetry often requires
the pair production of new particles at colliders and it
guarantees that cascade decays down to the lightest particle give
rise to missing energy plus jets and leptons. For a long time,
supersymmetry with the conserved R-parity was the only candidate
for such signals. However, any new physics with this type of new
symmetry may show up with similar signals and the discrimination
between different models at the LHC is quite challenging. In this
paper, we address the problem of discrimination between different
models, more concretely, in the little Higgs theory with T-parity
(LHT) and the supersymmetric theory with R-parity. We concentrate
on the pair production of heavy top partners, e.g., T-odd quarks
($T_-$) in LHT and the scalar top quarks ($\tilde{t}$) in the MSSM
at linear colliders (LC).
\end{abstract}

\preprint{ FERMILAB-PUB-07-054-T,
                  SNU TP 07-002 }

\

\maketitle

\section{Introduction}
Some of the most urgent theoretical issues in high energy physics,
the natural realization of electroweak symmetry breaking (EWSB)
and the study of its phenomenological implications at near future
experiments have drawn big interests among particle physicists.
Actually all the other sectors in the standard model of particle
physics (SM) other than the Higgs sector have been experimentally
well tested but the SM still has been regarded unnatural since the
mass of the Higgs boson is quadratically sensitive to the cutoff
scale ($\sim \Lambda_{\rm cut}$) which is supposed to be
hierarchically larger than the weak scale ($\sim M_W$).

Up to date, the low energy supersymmetry or its minimal
realization (MSSM) accommodating the SM in its particle spectrum
might have been regarded as the best motivated and successful
example of a natural theory of EWSB. Even better, with the exact
conservation of a discrete symmetry, dubbed R-parity, the lightest
supersymmetric particle (LSP) in the MSSM can be a nice candidate
of dark matter. The LSP (often the lightest neutralino
($\tilde{\chi}_0^1$)) appears in cascade decays of a
super-particles but easily escapes the particle detectors thus the
frequent appearance of missing energy signals can be thought as a
genuine feature of the MSSM with R-parity. Very interestingly,
this feature (the existence of a global symmetry and the stability
of the lightest particle in the newly extended sector) is shared
by recently suggested models beyond the SM. Among others, we are
mainly interested in little Higgs theories
~\cite{Arkani-Hamed:2001nc,Arkani-Hamed:2002qy}(for reviews, see~
\cite{Schmaltz:2005ky,Perelstein:2005ka}.) with a parity, dubbed
T-parity (LHT), by which the lightest new particle, a heavy photon
($A_H$) , is absolutely stable and any new particles in the
extended sector beyond the SM can be produced only by a pair.
Single production of the new particle is prohibited by the
T-parity just like the R-parity in the MSSM.

In LHT, the Higgs field is a Pseudo-Goldstone boson. Higgs
as a Goldstone boson has been considered since long time ago but
its natural realization in the light of collective symmetry
breaking was suggested quite recently. Once the Higgs mass is
protected by several symmetries, as was the case in the little
Higgs theories by construction, no one-loop quadratic mass could
be induced. However the original formulation of little Higgs
theories are severely constrained by electroweak precision data
~\cite{Csaki:2002qg,Csaki:2003si,Hewett:2002px}. This mainly came
from the fact that the newly introduced particles ($m \sim f$) in
the extended gauge and fermion sectors could directly mix with the
standard model particles at the tree level without suitable
symmetry protection. The most economic and elegant way out, up to
date, is to introduce T-parity, in the theory and assign the most
of new particles odd and the standard model particles even under
the parity~\cite{Cheng:2003ju,Cheng:2004yc,Low:2004xc}. By doing
so, no tree level mixing is allowed and electroweak constraints
are greatly relieved \cite{Hubisz:2005tx}:
\begin{equation}
f \gtrsim 500 {\rm GeV}.
\end{equation}
Once T-parity is introduced, the phenomenology of little Higgs
models could be essentially similar to the low energy
supersymmetry with R-parity \cite{Hubisz:2004ft}. The lightest
T-odd particle cannot further decay to the single standard model
particle similarly to the lightest neutralino $\tilde{\chi}_0^1$
in supersymmetric models. Because of this similarity, little Higgs
models can ``fake" the supersymmetric signals at the near future
colliders, such as the CERN Large Hadron Collider (LHC). Future
linear colliders, such as ILC and TESLA, could help to provide
valuable distinctions between MSSM and the other models, e.g.,
LHT. That's the main motivation of this study.

More specifically, we would concentrate on the phenomenology of
T-odd, new top quark (T) which is the lightest among all the newly
introduced fermions, in most realization of little Higgs models
with T-parity. The decay signal of T-quark (${\rm T} \rightarrow t
A_H$) can fake the signal of the scalar top quark
($\tilde{t}\rightarrow t \tilde{\chi}_0^1$). Note that provided
the mass of the scalar top is heavy enough ($m_{\tilde{t}}> m_t +
m_{\tilde{\chi}_0^1}$), scalar top can dominantly decay to the
standard model top quark and the neutralino.

This paper is organized as follows. In Sec.II, we setup the model
of little Higgs with T-parity. The mass and the gauge couplings of
T-quark are specified and the relevant Feynman rules are derived.
In Sec.III, we study the T-quark pair production by
electron-positron collision ($e^- e^+ \rightarrow T \bar{T}$). We
first calculate the total cross section of T-quark production and
compare it with the one of the scalar top pair production when
their masses are set to be the same. The angular distributions for
produced leptons taking the cascade decay of top quark is found to
give a clear distinction between the cases with T-quark (fermion)
in LHT and scalar top quark in the MSSM. The use of the
polarization of initial electron and positron is also discussed.
Summary will be given in the last section.
\section{Set-up: Top quark sector of LHT}
In this section, we will set up the top quark sector of little
Higgs model with T-parity. We restrict ourselves to the
$SU(5)/SO(5)$ realization of little Higgs mechanism to be specific
\cite{Arkani-Hamed:2002qy}. There have been lots of
phenomenological studies of the littlest Higgs models without
~\cite{Burdman:2002ns,Han:2003wu,Perelstein:2003wd,Park:2004ab} and with
T-parity~\cite{Birkedal:2006fz,Hubisz:2005bd,Blanke:2006sb,Chen:2006ie,Blanke:2006xr,Freitas:2006vy,Belyaev:2006jh,Carena:2006jx,Meade:2006dw,Matsumoto:2006ws,Choudhury:2006mp}.

To cancel out the 1-loop contribution of the standard model top
quark, the Yukawa sector of the third generation need to be
extended. To incorporate the collective symmetry breaking pattern,
the third generation quarks should be elevated to the complete
$SU(3)$ representations:
\begin{eqnarray}
Q_1 = \left(%
\begin{array}{c}
  q_1 \\
  T_1 \\
  0 \\
\end{array} \right), ~~Q_2 = \left(%
\begin{array}{c}
  0 \\
  T_2 \\
  q_2 \\
\end{array}%
\right),
\end{eqnarray}
where $q_1 \sim \mathbf{2}_{1/30}\times \mathbf{1}_{2/15}$, $q_2
\sim \mathbf{1}_{2/15}\times \mathbf{2}_{1/30}$ are doubles under
$[SU(2)\times U(1)]_{1}\times [SU(2)\times U(1)]_{2}$ gauge
symmetry of $SU(5)$ subgroup and $T_1 \sim \mathbf{1}_{8/15}\times
\mathbf{1}_{2/15}$ and $T_2 \sim \mathbf{1}_{2/15}\times
\mathbf{1}_{8/15}$ are singlets. Under T-parity operation, $Q_1
\rightarrow -\Sigma_0 Q_2$ where $\Sigma_0$ is the vev of an
$SU(5)$ symmetric tensor by which $SU(5)$ global symmetry is
broken down to $SO(5)$ at the energy scale $\sim f$. On top of the
standard model top quark, there are new T-even and T-odd quarks:
\begin{eqnarray}
T_\pm = \frac{1}{\sqrt{2}}\left(T_1 \mp T_2\right) \, ,
\end{eqnarray}
and their masses are
\begin{eqnarray}
M_{T_+} = \sqrt{\lambda_1^2+\lambda_2^2} f, ~~ M_{T_-} = \lambda_2 f \, ,
\end{eqnarray}
where $\lambda_i\simeq 1$ parameters are introduced to give the
Yukawa couplings. We can immediately notice that T-odd quark is
always lighter than T-even quark and its production would be
important at the near future colliders. The mass of standard model
top quark is given by
\begin{eqnarray}
M_t = \frac{\lambda_1 \lambda_2
v}{\sqrt{\lambda_1^2+\lambda_2^2}} \, ,
\label{topmass}
\end{eqnarray}
where $v \simeq 246 {\rm GeV}$ is the measured vacuum expectation value of the Higgs.
The gauge interaction with $Z$ boson and photon ($\gamma$) are immediately
read out. Because T's are $SU(2)$ singlets, there is no coupling
with $W$ boson.
\begin{eqnarray}
{\cal L} = \bar{T}_- \left[\frac{2}{3}\gamma_\mu \left( e A^\mu -
\frac{g}{c_w}s_w^2 Z^\mu \right)\right]T_- \, .
\end{eqnarray}
\begin{figure}[t]
\centerline{
  \includegraphics[width=.49\linewidth]{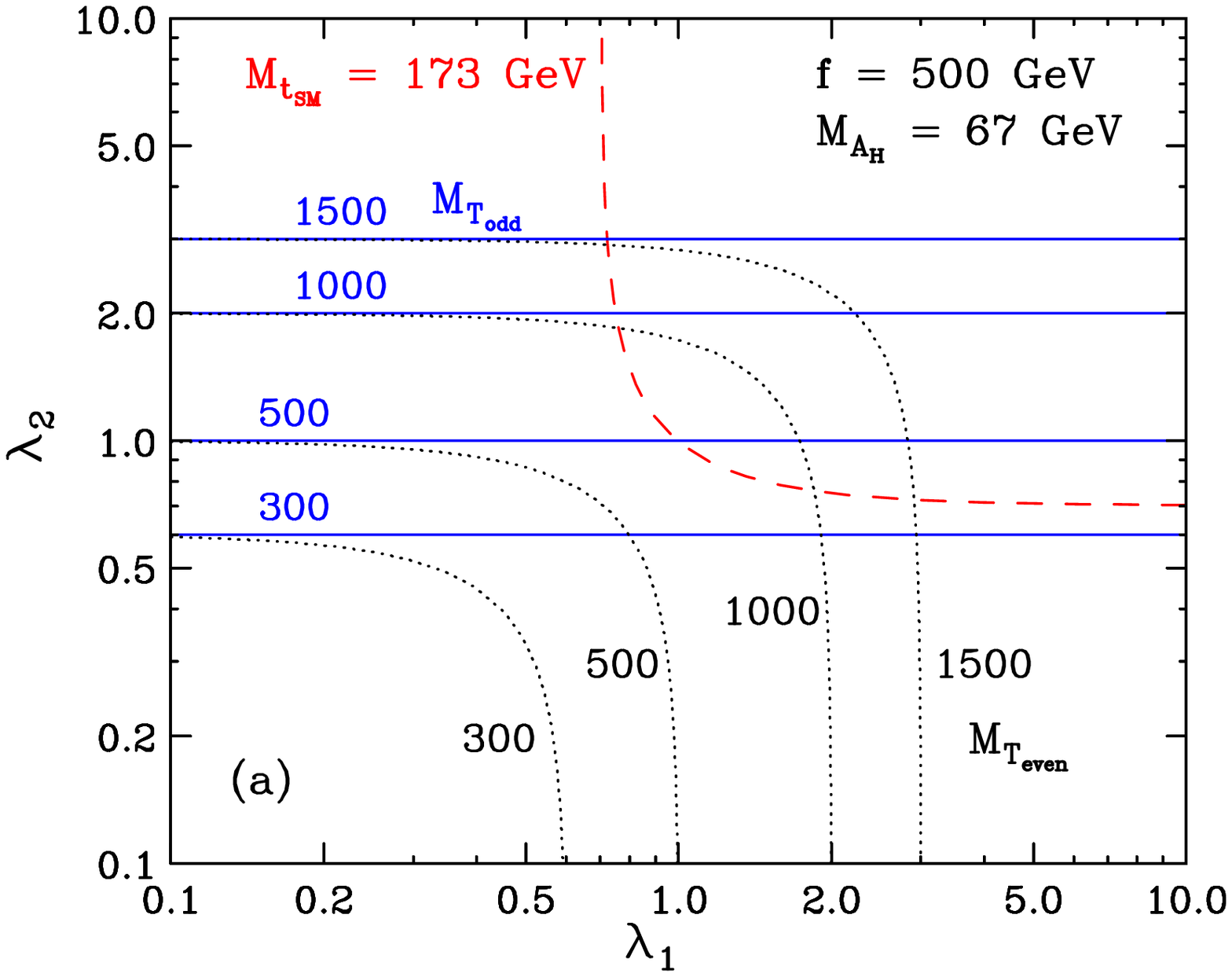}
   \includegraphics[width=.49\linewidth]{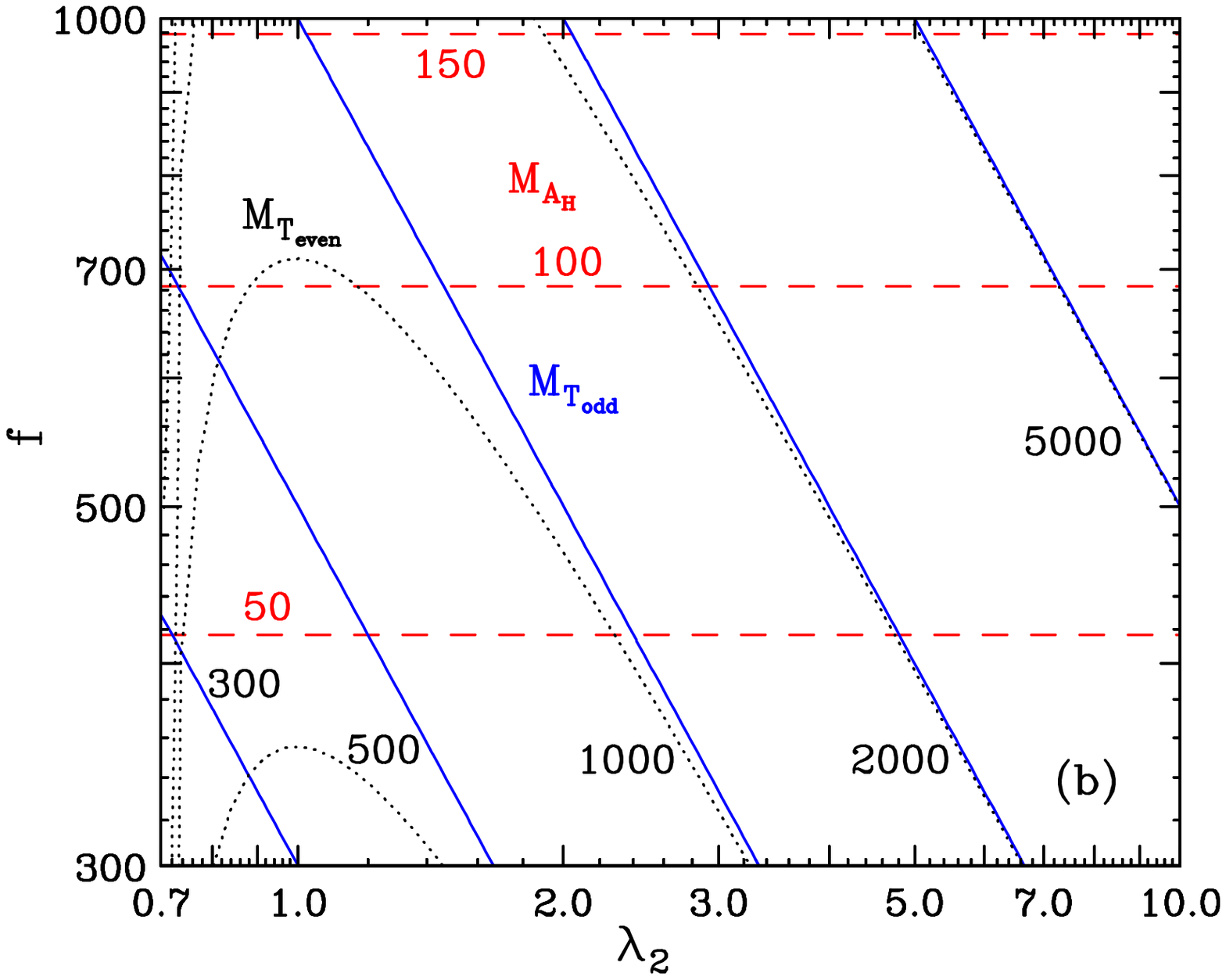} }
  \caption{The masses of T-even quark (black, dotted) and
T-odd quark (blue, solid) for given (a) $f=500$ GeV and
 (b) $M_{t_{SM}}=173$ GeV. In (a), a red dashed line shows a contour with
$M_{t_{SM}}=173$ in the plane of $\lambda_2$ versus $\lambda_11$.
In (b), $\lambda_1$ is determined assuming $M_{t_{SM}}=173$ GeV.}
  \label{lambda12}
\end{figure}
T-odd quark ($T_-$) could decay to the standard model top quark
and newly introduced neutral gauge bosons, $Z_H$ or $A_H$ through
\begin{eqnarray}
{\cal L}= \bar{T}_- \left[-\frac{2 g'c_\lambda}{5}\gamma_\mu
\left((P_R + c_\lambda \frac{v}{f}P_L)A_H^\mu + x_h
\frac{v^2}{f^2}P_R Z_H^\mu\right)\right]t  \, ,
\end{eqnarray}
where the mixing parameters $c_\lambda$ and $x_h$ are defined as
\cite{Hubisz:2004ft}:
\begin{eqnarray}
c_\lambda &\equiv&
\frac{\lambda_1}{\sqrt{\lambda_1^2+\lambda_2^2}} \, , \\  
x_h &\equiv& \frac{5 g g'}{4 (5g^2-g'^2)} \, .
\end{eqnarray}
Since $Z_H$ acquires heavy mass ($M_{Z_H}^2= g^2 (f^2 - v^2/4) $)
but the other one $A_H$ does not, ($M_{A_H}^2= g'^2 (f^2/5-v^2/4)
\simeq 0.16^2 f^2$), $T_-\rightarrow t A_H$ dominates the decay or
$BR(T_- \rightarrow t A_H) \simeq 1$ is a good approximation.

We show the masses of T-odd and T-even particles in Fig.~\ref{lambda12}.
We fix $f=500$ GeV and $M_{t_{SM}}=173$ GeV in Fig.~\ref{lambda12}(a).
Then the mass of the lightest T-odd particle (LTP)
is $M_{A_H}=67$ GeV. A red dashed contour line shows SM top mass in the plane of
$\lambda_2$ versus $\lambda_1$ and
blue solid lines represent masses of T-odd top partner for $M_{T_{odd}}=300, 500,
1000, 1500$ GeV. Black dotted lines represent the masses for the T-even
partners for $M_{T_{even}}=300, 500, 1000, 1500$ GeV.
However $\lambda_1$ is not an independent parameter if we
assume top quark mass.
From Eqn.~\ref{topmass}, it is determined by $\lambda_2$ and $M_T$ as follows,
\begin{eqnarray}
\lambda_1 &=& \frac{\lambda_2}{\sqrt{\lambda_2^2 \frac{v^2}{M_t^2}-1}} \, ,\\
\lambda_2 &>& \frac{M_t}{v} \sim 0.7 \, ,\\
f         &>& \sqrt{\frac{5}{4}} v \sim 275~{\rm GeV} \, .
\end{eqnarray}
Fig.~\ref{lambda12}(b) shows $M_{T_{odd}}$ in the plane of f versus
$\lambda_2$ assuming $\lambda_1$ is determined. $0.7 < \lambda_2 <
1$ and $400$ GeV $<f<$ $700$ GeV where $M_{T_{odd}}< 500$ GeV is
an interesting region for the pair production of T-odd partner
at the ILC with $\sqrt{s} \le 1$ TeV. Note that precision
electroweak constraints on this model consistently allows values
for $f$ as low as $500$ GeV without fine tuning~\cite{Hubisz:2004ft}.

We would add some comments about possible signals of T-even top partner
at the ILC and LHC. At the LHC, a single production of the T-even top partner
and a quark-jet is possible through T-channel exchange diagram of W-boson
\cite{Han:2003wu,Perelstein:2003wd}. Unfortunately the same production
mechanism does not work at the ILC because the T-even top partner does not
directly couple with electron and W-boson. There remains
a possibility that T-even top partners to be produced by pair through
S-channel photon and Z-boson exchange diagrams. But it is highly
suppressed or disallowed by kinematics in most of parameter space we are
considering. For details, see the Appendix-C.
Considering all these issues, we would like to concentrate on phenomenology of T-parity odd  partners ($T_-$).
It is also fair to assume that the top partner is odd under a discrete symmetry since we plan to compare with stop in SUSY.
We will use a study point, $\lambda_2 = 0.8$ and $f=500$ GeV, for which pair production of T-even partners is not allowed at ILC with $\sqrt{s} = 1 {\rm TeV}$. Our analysis can be also applied to fermionic partner of top in any models with a discrete symmetry such as KK-top in UED model.

\section{T-odd top pair production at LC}
\subsection{Production and decay}
\begin{figure}[t]
\centerline{
  \includegraphics[width=.28\linewidth]{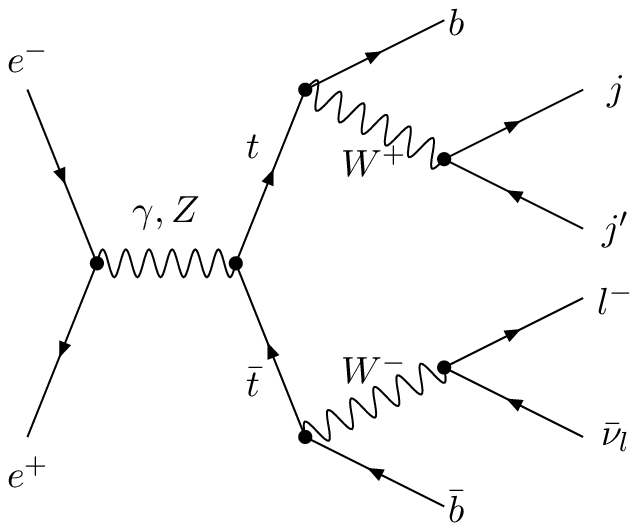}   
  \hspace{0.5cm}\includegraphics[width=.3\linewidth]{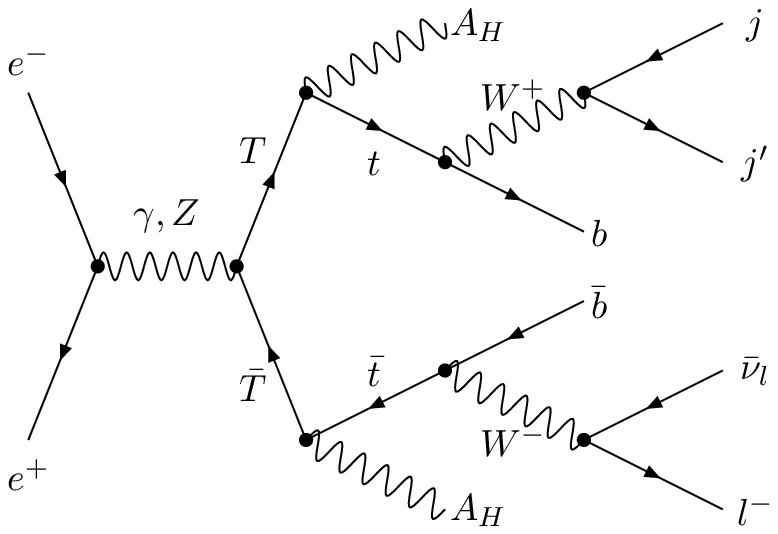}
  \hspace{0.5cm}\includegraphics[width=.3\linewidth]{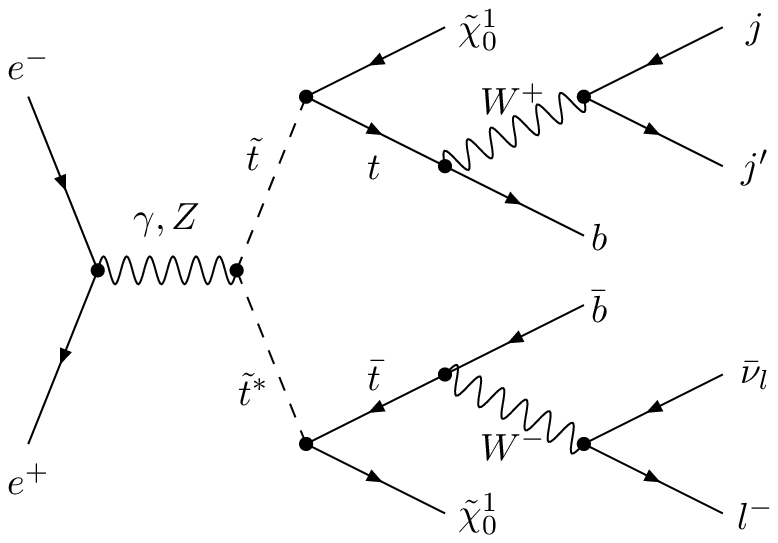} }
  \caption{Feynman diagrams for pair productions of top quark in the SM,
T-odd quarks/scalar top quarks
  and their subsequent decay to $tA_H$/$t \tilde{\chi}_1^0$ in LHT and MSSM, respectively.
The final state signals are identical ($l^- j j b b +\met$).}
  \label{feynman}
\end{figure}

The top quark pair production mechanisms (with or without the missing energy) in the SM, LHT and MSSM are depicted in Fig.~\ref{feynman},

The pair production cross section of T-odd fermion is calculated
based on the interaction given in section 2.
\begin{eqnarray}
\frac{d\sigma}{d\cos\theta} &=& \frac{N_c \beta s}{128 \pi} \left
[{\cal A}\left(1+\beta\cos\theta \right )^2+{\cal B}\left (
1-\beta\cos\theta \right )^2 +{\cal C} \left ( 1-\beta^2 \right
)\right],\\
\sigma &=& \frac{N_c \beta s}{64\pi} \left [ ({\cal A+B})\left (
1+ \frac{1}{3}\beta^2 \right ) + {\cal C} \left ( 1-\beta^2
\right) \right],
\end{eqnarray}
where $N_c=3$ is the number of color charges, $\beta = \sqrt{1-
\frac{4m^2}{s}}$ is the velocity of the final state fermion and
the effective product of couplings and propagator factors are
conveniently defined as
\begin{eqnarray}
{\cal A}=  \left ( |G_{LL}|^2 + |G_{RR}|^2 \right ), {\cal
B}=\left ( |G_{LR}|^2 + |G_{RL}|^2 \right ), {\cal C}= 2 {Re}
\left ( G_{LL} G_{LR}^*+G_{RR}G^*_{RL} \right )
\end{eqnarray}
where $G_{AB}$ ($A$ and $B$ denote Left and Right-handed chiralities.) are given as
\begin{equation}
G_{AB}(s) = \sum_X \frac{g_A(X\to\ell\bar{\ell}) g_B(X\to
f\bar{f})}{s- M_X^2+iM_X\Gamma_X} \, ,
\end{equation}
summed over all contributing gauge bosons, in this case $X=\gamma,
Z$ (see~\cite{Barger:1987nn} for detail.).
$\gamma$ contribution is dominant and $Z$ contribution is
roughly 10\% of the total cross section. The cross term is
negligible (smaller by two order of magnitude). The cross section
is calculated with {\tt CalcHEP}~\cite{Pukhov:2004ca} and
cross-checked with {\tt
MadGraph/MadEvent}~\cite{Stelzer:1994ta,Maltoni:2002qb}. One important
point we should notice here is that because T-quark is a
vector-like fermion, $G_{LL} = G_{LR}$ and $G_{RR} = G_{RL}$:
\begin{eqnarray}
G_{LL} &=& G_{LR} = \frac{(-e)(\frac{2}{3}e)}{s} + \frac{g_L g_{ZT\bar{T}}}{s-M_Z^2} \, , \\
G_{RR} &=& G_{RL} = \frac{(-e)(\frac{2}{3}e)}{s} + \frac{g_R
g_{ZT\bar{T}}}{s-M_Z^2} \, .
\end{eqnarray}
Here $g_L$ and $g_R$ are the couplings between $Z$ and $e^+e^-$
given as $g_{L/R} = \frac{e}{s_W c_W} \big ( I_3 - Q_e s_W^2 \big
)$ where $I_3=-\frac{1}{2}$ is the third component of weak
iso-spin of the left-chiral electron, $Q_e=-1$ its charge and the
coupling between $Z$ and $T\bar{T}$ is given as $g_{ZT\bar{T}} =
-\frac{2}{3} \frac{g}{c_W}s_W^2$. As a consequence, we get a
simple relation between ${\cal A,B}$ and ${\cal C}$:
\begin{eqnarray}
{\cal A = B} = \frac{1}{2}{\cal C}.
\end{eqnarray}
What can we learn from this observation? First of all, as is
explicitly shown in the Appendix, the forward-backward asymmetry
vanishes. This is actually a generic feature of a vector-like
fermion. Based on the precise measurement on $A_{FB}$ asymmetry,
one will be able to clearly prove that the produced T-quark is a
vector-like fermion even though its direct measurement might be
challenging. Furthermore, the Left-Right asymmetry, $A_{LR}$ can provide a useful
information about the couplings between T-quark and vector bosons
$\gamma$ and $Z$ (see Appendix).

In Fig.~\ref{xsection}(a), we show the pair production cross section of
T-odd partner as a function of $f$ at a 1 TeV linear collider for
various values of $\lambda_2$.
As we mentioned earlier, photon dominantly contributes to total cross section.
However, $Z$ contribution also becomes important when polarized beams are used and
Fig.~\ref{xsection}(b) shows production cross section with polarized beams
for $f=500$ GeV and $\lambda_2=0.8$ ($M_{T_{odd}}=400$ GeV).
Polarized beams may be used to confirm
the nature of T-odd partner by measuring cross sections with different polarizations of the beams.
As $\lambda_2$ increases, the
corresponding cross section decreases since the mass of T-odd
partner is proportional to $\lambda_2$. The pair
production of T-odd partner with $f \ge 700$ GeV at such a collider
is not allowed as shown in Fig.~\ref{lambda12}(b).
Solid lines represent cross section without the
initial state radiation (ISR) while the dotted lines are ISR
corrected.
\begin{figure}[t]
\centerline{
  \includegraphics[width=.49\linewidth]{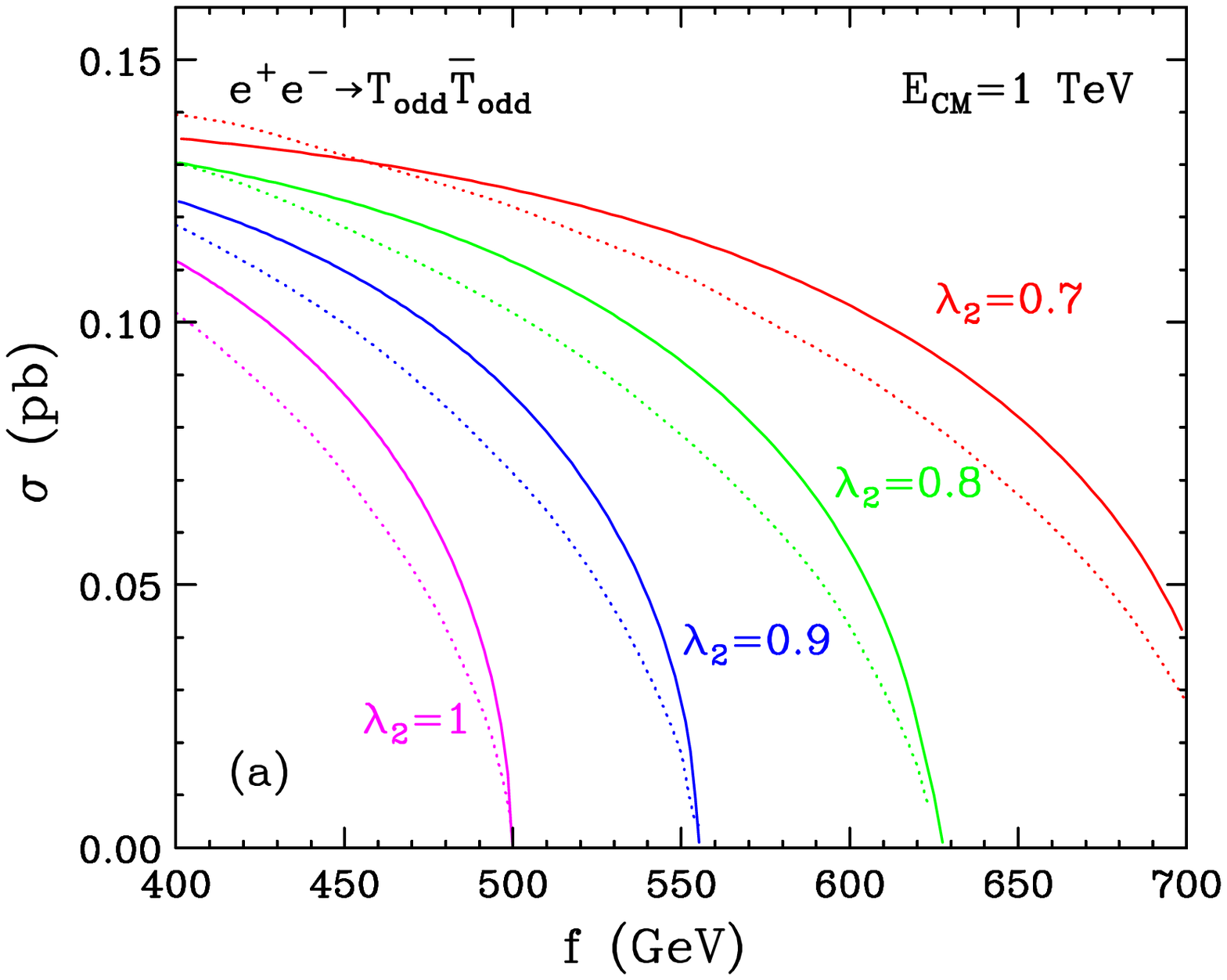}
  \includegraphics[width=.48\linewidth]{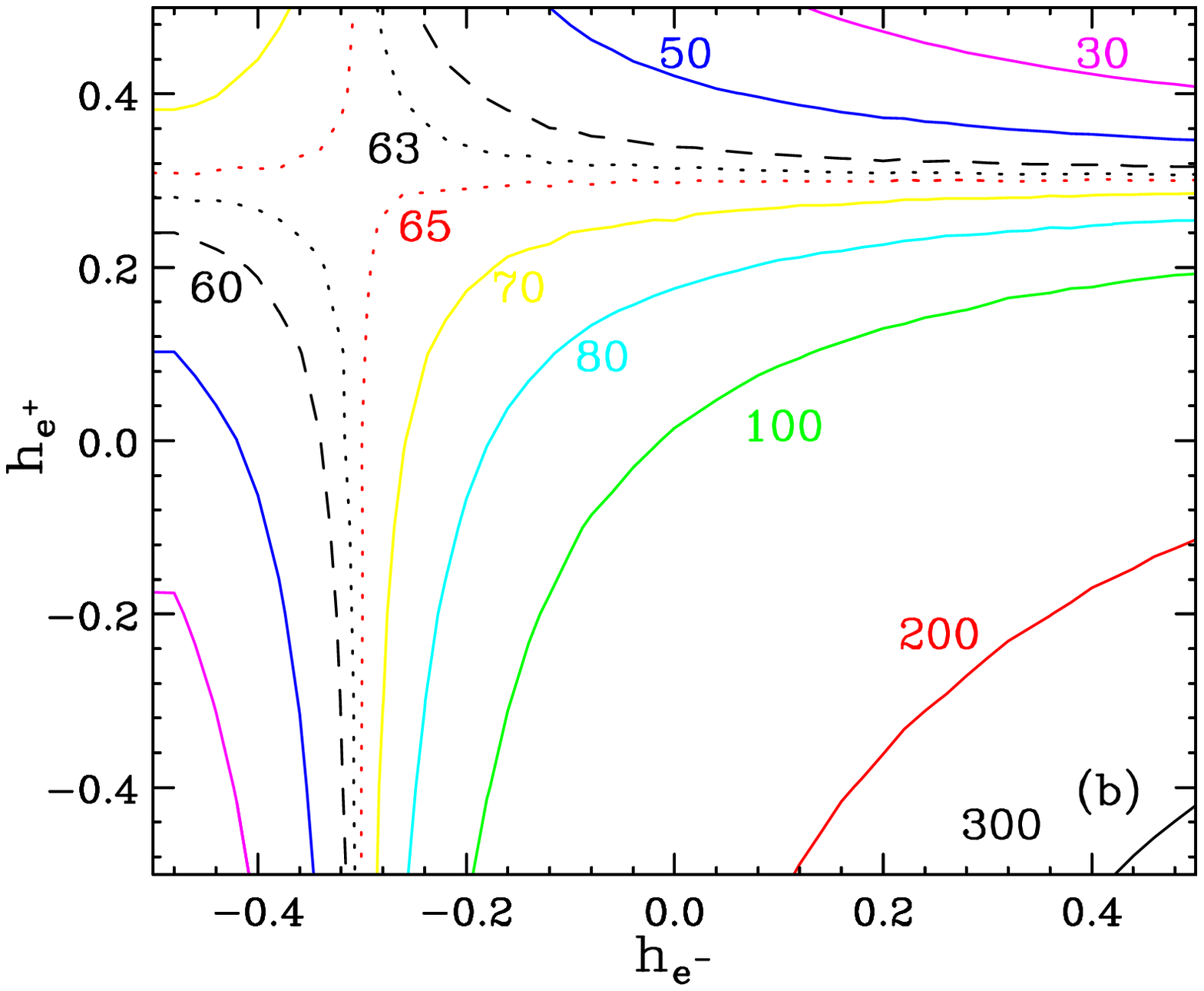} }
  \caption{Cross sections for T-odd quark pair production
(a) with unpolarized beam for various values of $\lambda_2$ in pb and
(b) with the initial helicity states of electron and positron in fb.
The dotted lines in (a) are ISR-corrected.}
  \label{xsection}
\end{figure}
Generally the cross section at the linear collider is given by
\begin{eqnarray}
\sigma \left ( e^+e^- \to T\bar{T} \right ) &=& \int dx_1 dx_2
\left [ f_{e/e}(x_1) f_{e/e}(x_2) \hat\sigma \left ( e^+e^- \to T\bar{T} \right ) \right . \nonumber \\
\hspace{1cm}&+& \left . \left (f^{brem}_{\gamma/e}(x_1) +
f^{beam}_{\gamma/e} (x_1) \right )
  \left ( f^{brem}_{\gamma/e}(x_2) + f^{beam}_{\gamma/e} (x_2) \right )
  \hat\sigma ( \gamma\gamma \to T\bar{T} )
 \right ] \\
&\approx& \int dx_1 dx_2 \left [ f^{brem}_{e/e}(x_1)
f^{brem}_{e/e}(x_2) \hat\sigma \left ( e^+e^- \to T\bar{T} \right
) \right ] \, .
\end{eqnarray}
Here
$f_{\gamma/e}(x)$ is the parton distribution function for finding
a photon inside the electron beam and
$f_{e/e}(x)$ is the parton distribution function for finding
an electron inside the electron beam. $f_{e/e}(x)$ is given by the
convolution of bremstrahlung (or ISR) and beamstrahlung,
\begin{equation}
f_{e/e}(x) = \int_x^1 \frac{dz}{z} f^{brem}_e \left
(\frac{x}{z}\right ) f^{beam}_e(z) \, ,
\end{equation}
where $f^{beam}_e(x)$ is the beamstrahlung distribution function
of the electron and $f^{brem}_e$ is the bremstrahlung distribution
function of the electron.
{\tt CalcHEP}~\cite{Pukhov:2004ca} realizes the bremstrahlung function
with the expression in~\cite{Jadach:1988gb,Skrzypek:1990qs},
\begin{equation}
f^{brem}_e (x) = e^{\beta (3/4-\gamma)}\beta(1-x)^{\beta-1}
\frac{(1+x^2)-\beta ((1+3x^2)\log(x)/2 + (1-x)^2)/2}{
2\Gamma(1+\beta)} \, ,
\end{equation}
where $\beta=\frac{\alpha}{\pi} \left ( 2 \log(\frac{Q}{m_e})-1
\right )$, $\gamma=0.5772156649$ is the Euler constant, $\Gamma$ is
the gamma function, $m_e$ is the electron mass, and
$\alpha=1/137.0359895$ is the fine structure constant. We take ISR
scale, $Q= \sqrt{s}$.
We find that in our study the beamstrahlung effect (for the small
beamstrahlung parameter) in the cross sections and the
contribution from $\sigma ( \gamma\gamma \to T\bar{T} )$ are small
enough to ignore (See~\cite{Datta:2005gm,Accomando:2004sz}
for the beamstrahlung effect
at the high energy linear colliders such as CLIC).
In Fig.~\ref{xsection}(a), we can notice that the inclusion of ISR gives smaller cross
sections (except for $\lambda_2=0.7$) since the beam loses its
energy. However for the case of the light T-odd partner (i.e.,
$\lambda_2 \sim 0.7$ and $f<500$ GeV), the ISR increases cross
sections since the energy loss of the beam results in getting
closer to the pair production threshold which behaves as
$\sigma_{LH}\sim\beta$ where $\beta = \sqrt{1-\frac{4
M^2_{T_{odd}}}{s}}$. In the case of pair production of scalar
partners (i.e., stop), the threshold rises as
$\sigma_{SUSY}\sim\beta^3$.
The production of fermion (T-odd partner) and the production
of scalar (stop) have different behavior near the threshold as shown in Fig.~\ref{threshold}.
\begin{figure}[t]
\centerline{
  \includegraphics[width=.6\linewidth]{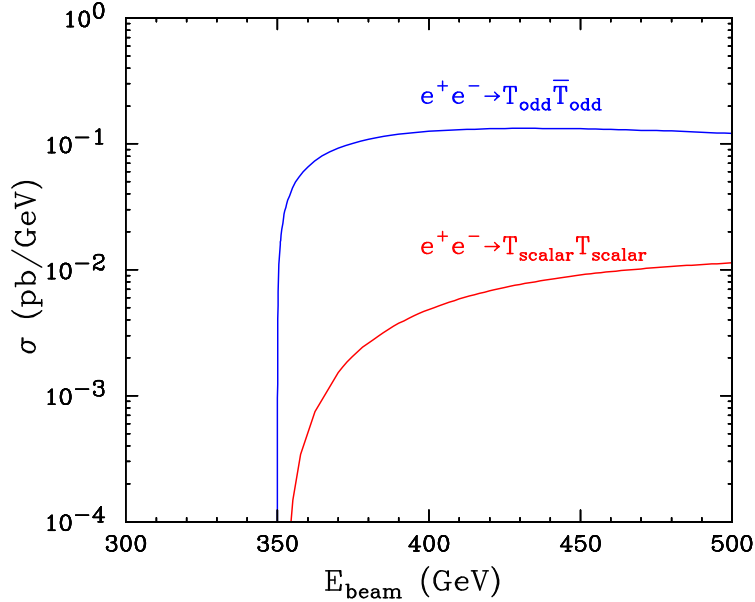} }
  \caption{Pair production cross section in LHT (blue) and SUSY (red)
near the threshold}
  \label{threshold}
\end{figure}
It is well known that in the massless limit of new particles, the
cross section ratio converges to a factor of 4 due to the fact
that there are two different helicity states in LH while in SUSY,
stop is a scalar particle~\cite{Battaglia:2005zf,Battaglia:2005ma,Choi:2006mr}.
Fermions prefer the forward or backward direction when they are produced
at $e^+e^-$ collider,
\begin{equation}
\left ( \frac{d\sigma}{d\cos\theta} \right )_{LH}   \sim  2 (1+\cos^2\theta) \, ,
\end{equation}
while scalar particles are produced more in the central region,
\begin{equation}
\left ( \frac{d\sigma}{d\cos\theta} \right )_{SUSY} \sim
(1-\cos^2\theta) \, .
\end{equation}
These particles go through further decays and
we consider semi-leptonic decay of SM top.
We assume that we know the mass and momentum of $W$ and $t$ from two jets and one b-jet.
Fig.~\ref{angular distribution} shows angular distributions in LH and MSSM in
the semi-leptonic channel. The distribution for MSSM is similar to
the angular distribution of the scalar production. We see clear difference between two
distributions.
When top partners are produced, T-odd fermions prefer forward or backward direction while
stops are mainly produced in the central region, as explained above.
There is no forward-backward asymmetry at this production level.
Now they continue to decay into to SM top to $W_\mu$ and $b$.
As noticed in~\cite{Kane:1991bg}, in the case of SM top pair production at LC,
the forward-backward asymmetry appears due to huge interference between $\gamma$ and $Z$
(negative in backward and positive in forward region),
although there is no FB asymmetry at all in $\gamma$-mediation and small FB asymmetry appears
in the  $Z$-mediation.
This characteristic feature is carried in the decay of T-odd and
the angular distribution of top at LC in the production just looks like
angular distribution of the lepton in the case of LH.
However, in the case of SUSY, the angular distribution is not spoiled
as much as LH case since they are produced in the central region.
(See~\cite{Datta:2005zs,Meade:2006dw,Barr:2004ze,Wang:2006hk,Smillie:2005ar,Alves:2006df} for the
discrimination between SUSY and other new physics such as UED/LH at the LHC in a particular
cascade decay. These studies seem to either require large luminosity or depend on
certain assumptions.)
\begin{figure}[t]
\centerline{
  \includegraphics[width=.49\linewidth]{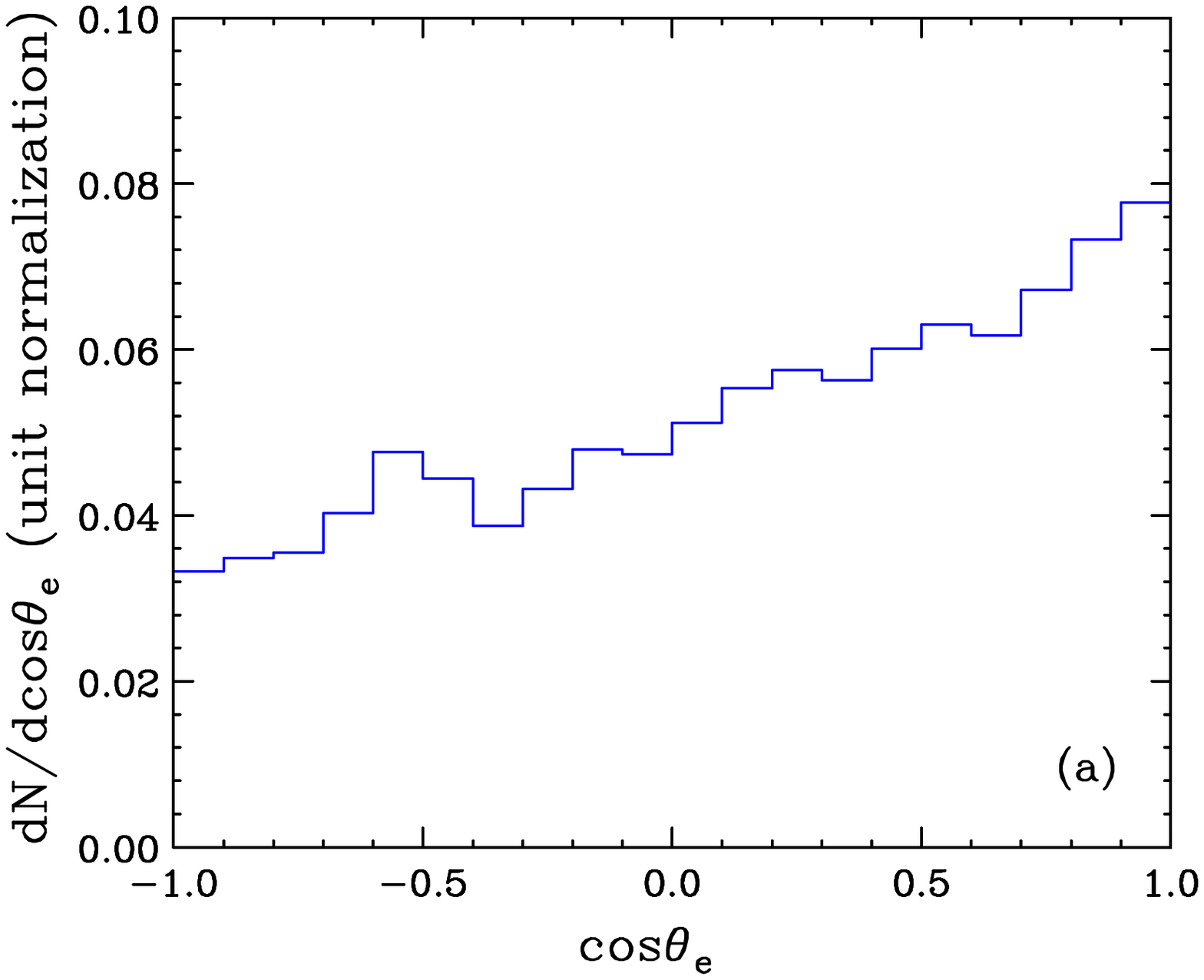}
  \includegraphics[width=.49\linewidth]{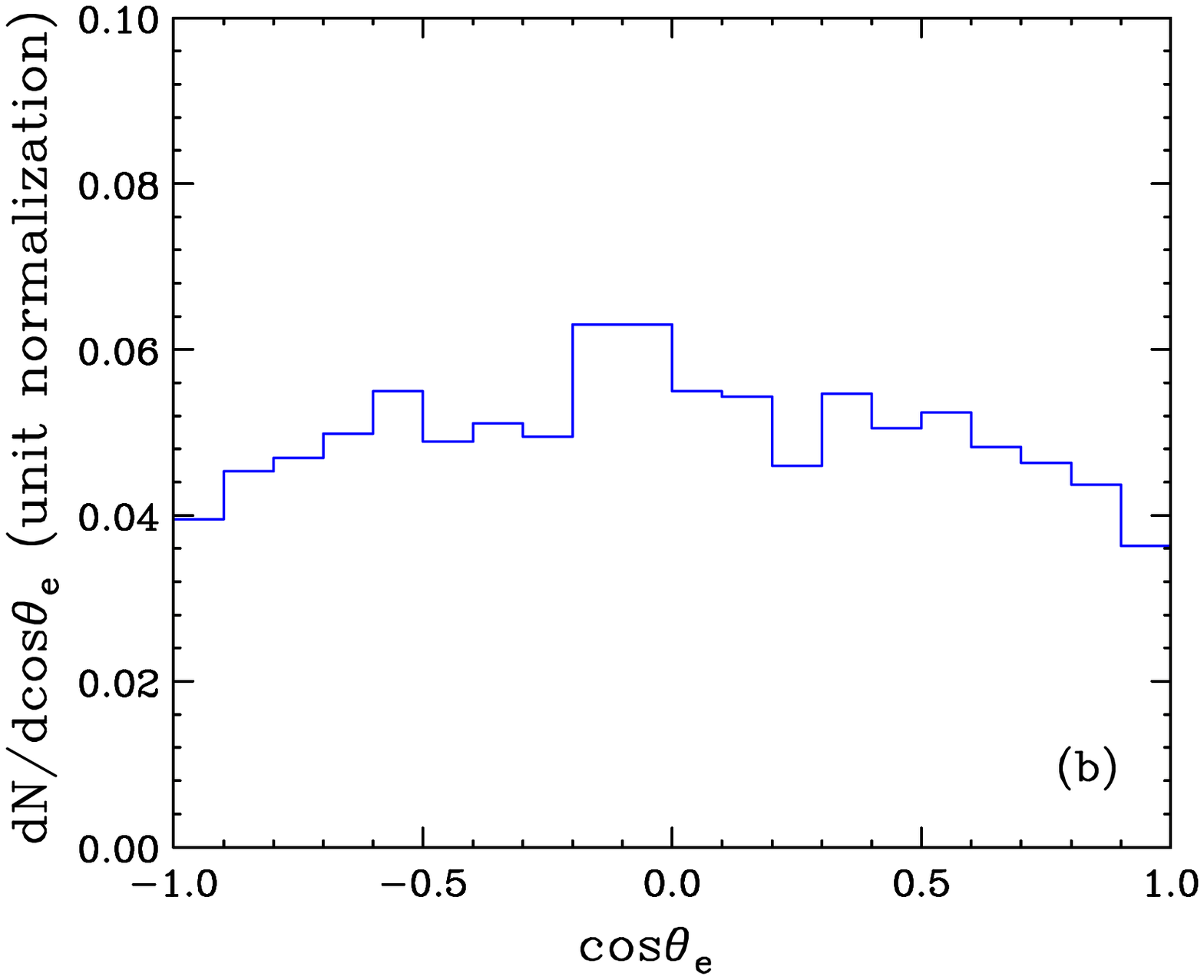} }
  \caption{The angular distribution of final state lepton ($\ell^-$) for
LHT and MSSM, respectively. In LHT, leptons follow the angular distribution
of the parent T-odd quarks and highly anisotropic.
However in MSSM, mainly because of the characteristics of
scalar particle production, the angular distribution is roughly $(1-\cos^2\theta)$.
({\tt PYTHIA}~\cite{Sjostrand:2000wi}
also gives similar distribution in the case of MSSM.)}
  \label{angular distribution}
\end{figure}
\subsection{Backgrounds}

We list possible backgrounds for our signal in Table~\ref{tab:background}.
The dominant backgrounds are $t\bar{t}$ and $\nu_e e^+W^-Z$.
$t\bar{t}$ can be easily distinguished from our signal in the energy distribution of SM top since
$E_{top} = \frac{\sqrt{s}}{2}$ in SM while LHT or SUSY shows a flat distribution over wide range of
$E_{top}$ as shown in Fig.~\ref{e_top}.
In fact, the upper edge ($E_+$) is always less than $E_{top} = \frac{\sqrt{s}}{2}$ and
signal is well separated from this background.
A background $t\bar{t} Z$ is included when $t\bar{t}\nu \bar{\nu}$ is estimated.
The last three backgrounds have two b-jets from $Z$ decay
and therefore they can be eliminated by imposing relevant cut on invariant mass of two b-jets.
The rest of background are much smaller than our signal cross section,
$\sigma(e^+e^- \to T\bar{T} \to 2j+2b+\ell +\met )=29$ fb assuming
${\cal O} \big ( \sigma(e^+e^- \to t\bar{t}) \big )\sim 100$ fb.
These backgrounds are also ignored in supersymmetric case (see~\cite{Kitano:2002ss} for details)
with similar reasons.
However our case is even better since the pair production cross section of two fermions (T-odd) is
much larger than the pair production cross section of two scalars (stop),
as discussed in the previous section.
With the signal cross section and the expected ILC luminosity ${\cal L}=500~{\rm or}~1000$ fb$^{-1}$,
we can easily get large number events
even if we lose some of events by imposing cuts to remove SM backgrounds.
Therefore we can safely ignore backgrounds in our study.
\begin{table}[t]
\centerline{
\begin{tabular}{|l|l|}
\hline
  $\sigma(e^+e^- \to t\bar{t})$ = 173
& $\sigma(e^+e^- \to t\bar{t} \to 2j+2b+\ell +\met )$ = 50 \\
  $\sigma(e^+e^- \to t\bar{t} \nu_e \bar{\nu}_e)$ = 0.74
& $\sigma(e^+e^- \to t\bar{t} \nu_e \bar{\nu}_e \to 2j+2b+\ell +\met )$ = 0.22 \\
  $\sigma(e^+e^- \to t\bar{t} \nu_\mu \bar{\nu}_\mu)$ = 0.317
& $\sigma(e^+e^- \to t\bar{t} \nu_\mu \bar{\nu}_\mu \to 2j+2b+\ell +\met )$ = 0.09 \\
  $\sigma(e^+e^- \to ZW^+W^- )$ = 56.8
& $\sigma(e^+e^- \to ZW^+W^- \to 2j+2b+\ell +\met )$ = 2.5 \\
  $\sigma(e^+e^- \to \nu_e e^+W^-Z   )$ = 165.45
& $\sigma(e^+e^- \to \nu_e e^+W^-Z   \to 2j+2b+\ell +\met )$ = 2.68 \\
  $\sigma(e^+e^- \to \nu_\mu \mu^+W^-Z   )$ = 6.3
& $\sigma(e^+e^- \to \nu_\mu \mu^+W^-Z   \to 2j+2b+\ell +\met )$ = 0.65 \\
\hline
\end{tabular}
}
\label{tab:background}
\caption{SM backgrounds (in fb) in semi-leptonic channel are estimated using {\tt MadEvent/MadGraph}
and cross-checked with {\tt CalcHEP}.
We used $BR(Z\to \nu\bar{\nu})=0.6666$,
$BR(t\to W^+b)=1$, $BR(W^\pm \to jj')=0.6796$, $BR(W^\pm\to\ell^\pm \nu_\ell)=0.1068$ and
$BR(Z\to b\bar{b})=0.1514$.}
\end{table}

\subsection{Energy distribution of reconstructed top}

\begin{figure}[t]
\centerline{
  \includegraphics[width=.49\linewidth]{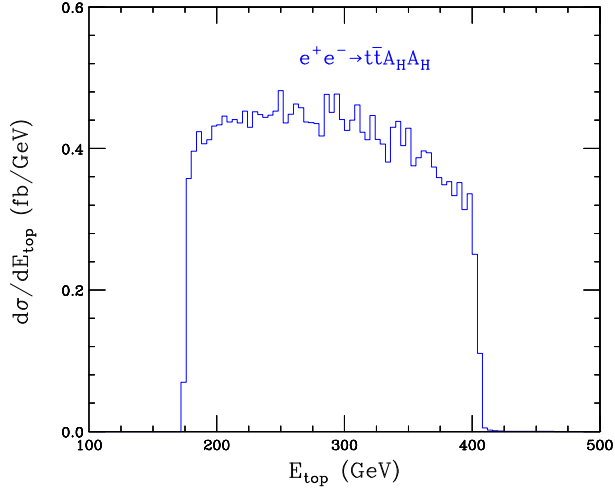} }
  \caption{The energy distribution of top quark}
  \label{e_top}
\end{figure}

In the hadronic (semi-leptonic channel), we can reconstruct momenta of two (one)
top quarks and look at the energy distribution of top. There are two
endpoints given by
\begin{eqnarray}
E_{+/-} = \frac{\gamma}{2 M_T} \left ( M_T^2 - M_N^2 + m_t^2 \pm
\beta \sqrt{ \big [ M_T^2 - \big ( M_N + m_t \big )^2 \, \big ]
             \big [ M_T^2 - \big ( M_N - m_t \big )^2 \, \big ] }  \right ) \, ,
\end{eqnarray}
where $\beta = \sqrt{1 - \frac{4 M_T^2}{s}}$, $\gamma = \frac{1}{\sqrt{1-\beta^2}}$,
$M_T$ denotes the mass for T-odd fermion or stop, $M_N$ for $A_H$ or $\tilde{\chi}_1^0$
and $m_t$ for SM top.
From the measurement of these two end points, we get two unknown masses
\begin{eqnarray}
M_T &=& \sqrt{s}\frac{\sqrt{E_+ E_-}}{E_-+E_+} \frac{1}{\sqrt{2}}
\sqrt{1+ \frac{m_t^2}{E_+ E_-} +
\sqrt{\left ( 1- \frac{m_t^2}{E_+^2} \right )
\left (1 - \frac{m_t^2}{E_-^2} \right )} } \, , \\
M_N &=& M_T \sqrt{1- \frac{2 \big ( E_+ + E_-\big )}{\sqrt{s}} + \frac{m_t^2}{M_T^2}} \, .
\end{eqnarray}
This method is usually discussed in the slepton pair production and
taking top quark mass to be zero, we recover well known formulas
\begin{eqnarray}
E_{+/-} &=& \frac{\sqrt{s}}{4} \left ( \frac{M_T^2-M_N^2}{m_T^2}
\right ) \left ( 1\pm \sqrt{1- \frac{4 M_N^2}{s}} \right ) \, , \\
M_T &=& \sqrt{s}\frac{\sqrt{E_+ E_-}}{E_-+E_+} \, , \\
M_N &=& M_T \sqrt{ 1- 2 \frac{E_+ + E_-}{\sqrt{s}}} \, .
\end{eqnarray}
The accuracy of this method depends
on how well we can reconstruct top quark momentum.
For our study point, $m_T = 400$ GeV and $m_{A_H} = 66.76$ GeV,
we get two endpoints at $E_- = 174$ GeV and $E_+ = 406$ GeV,
as shown in Fig.~\ref{e_top}.
The distribution gets smeared near $E_+$ due to ISR at LC.
\subsection{Angular distribution of b-jet}
In~\cite{Kitano:2002ss}, it was noticed that the information about the stop
and the LSP is imprinted in the helicity of the top quark in the stop decay,
and hence the measurement of the helicity of the top quark provides some
knowledge of these particles. We follow the same analysis given in~\cite{Kitano:2002ss}.
The Lagrangian is given by
\begin{equation}
{\cal L} = \bar{T}_- \gamma_\mu (f_L P_L + f_R P_R) A^\mu_H t + H.C. \, ,
\end{equation}
where $P_{L/R} = \frac{1\mp \gamma_5}{2}$.
Using the spin vector method,
the spin vector $S^\mu$ for the top quark in the decay process, $T\to t A_H$ is given by
\begin{eqnarray}
\frac{1}{2}N ( \rlap{\,/}P_t + m_t ) (1+ \rlap{\,/}S\gamma_5) &\equiv&
(\rlap{\,/}P_t+m_t)\gamma^\mu (f_L^\ast P_L + f_R^\ast P_R) (\rlap{\,/}P_T+M_T)\gamma^\nu
(f_L P_L + f_R P_R) (\rlap{\,/}P_t + m_t) \nonumber \\
&&\hspace{2cm}\times \left ( -g_{\mu\nu}+\frac{P_{A_H}^\mu P_{A_H}^\nu}{M_{A_H}^2} \right ) \, ,
\end{eqnarray}
with $P_t$, $P_{A_H}$ and $P_T$ being four-momenta of the top quark, the LTP and the T-odd top,
respectively. From this equation, $N$ and $S^\mu$ are obtained as follows.
\begin{eqnarray}
N &=& \left (|f_L|^2 + |f_R|^2 \right ) \left [
\big ( M_T^2 + m_t^2 - M_{A_H}^2 \big ) +
\frac{\big ( M_T^2 - m_t^2 + M_{A_H}^2  \big )
      \big ( M_T^2 - m_t^2 - M_{A_H}^2 \big )}{M_{A_H}^2}
 \right ] \nonumber \\
&&\hspace{3cm}- 12 m_t M_T Re \big ( f_L^\ast f_R \big ) \, , \\
N S^\mu &=& \frac{|f_L|^2 - |f_R|^2}{m_t M_{A_H}^2}
\big [
\big( (M_T^2-m_t^2 )^2 + 3 M_{A_H}^2 (M_T^2-m_t^2 ) -2M_{A_H}^4 \big) P_t^\mu \nonumber \\
&& \hspace{5cm}-2m_t^2 \big ( M_T^2 - m_t^2 + 2M_{A_H}^2\big ) P^\mu_{A_H} \big ] \, .
\end{eqnarray}
\begin{figure}[t]
\centerline{
  \includegraphics[width=.49\linewidth]{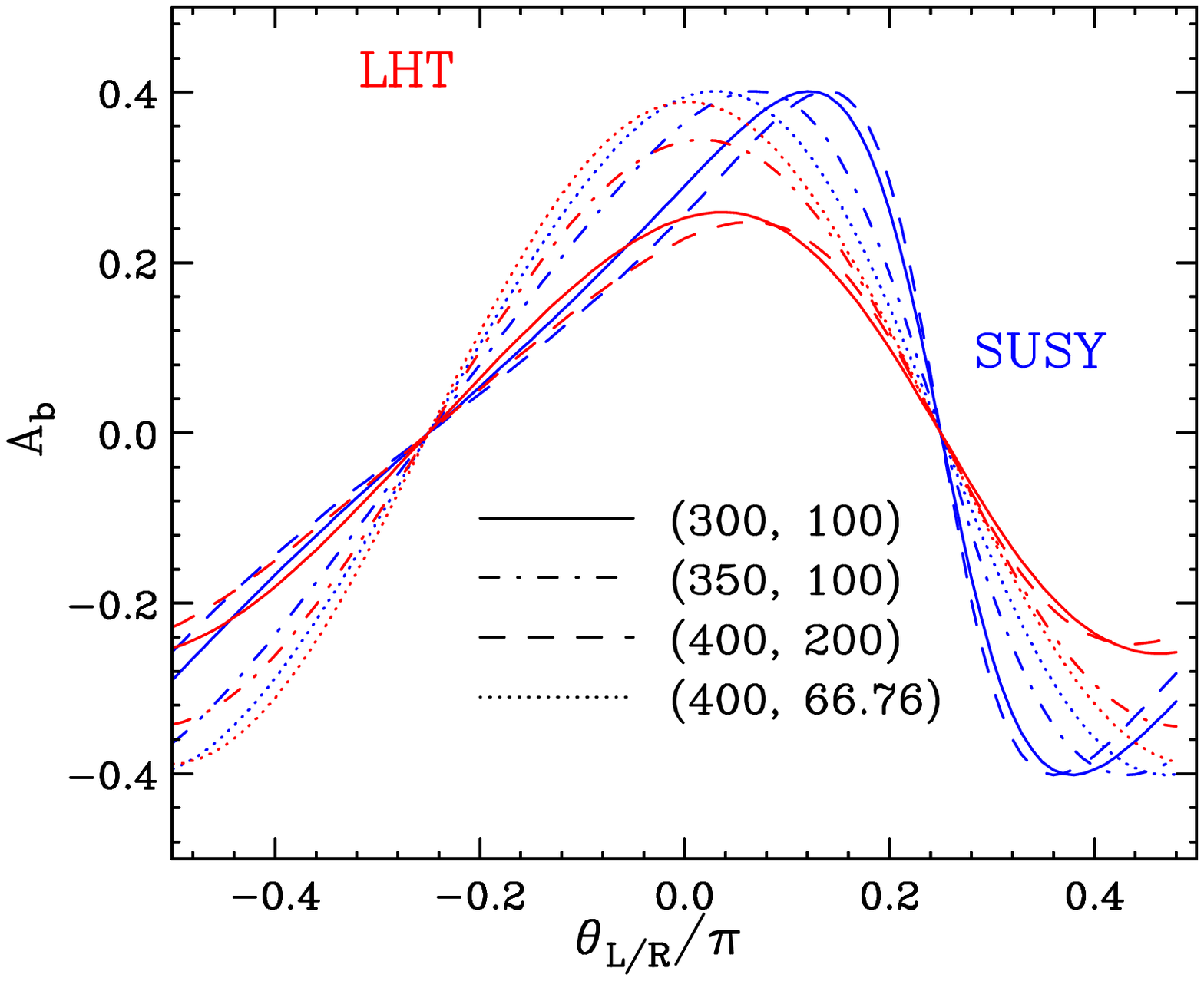}
  \includegraphics[width=.49\linewidth]{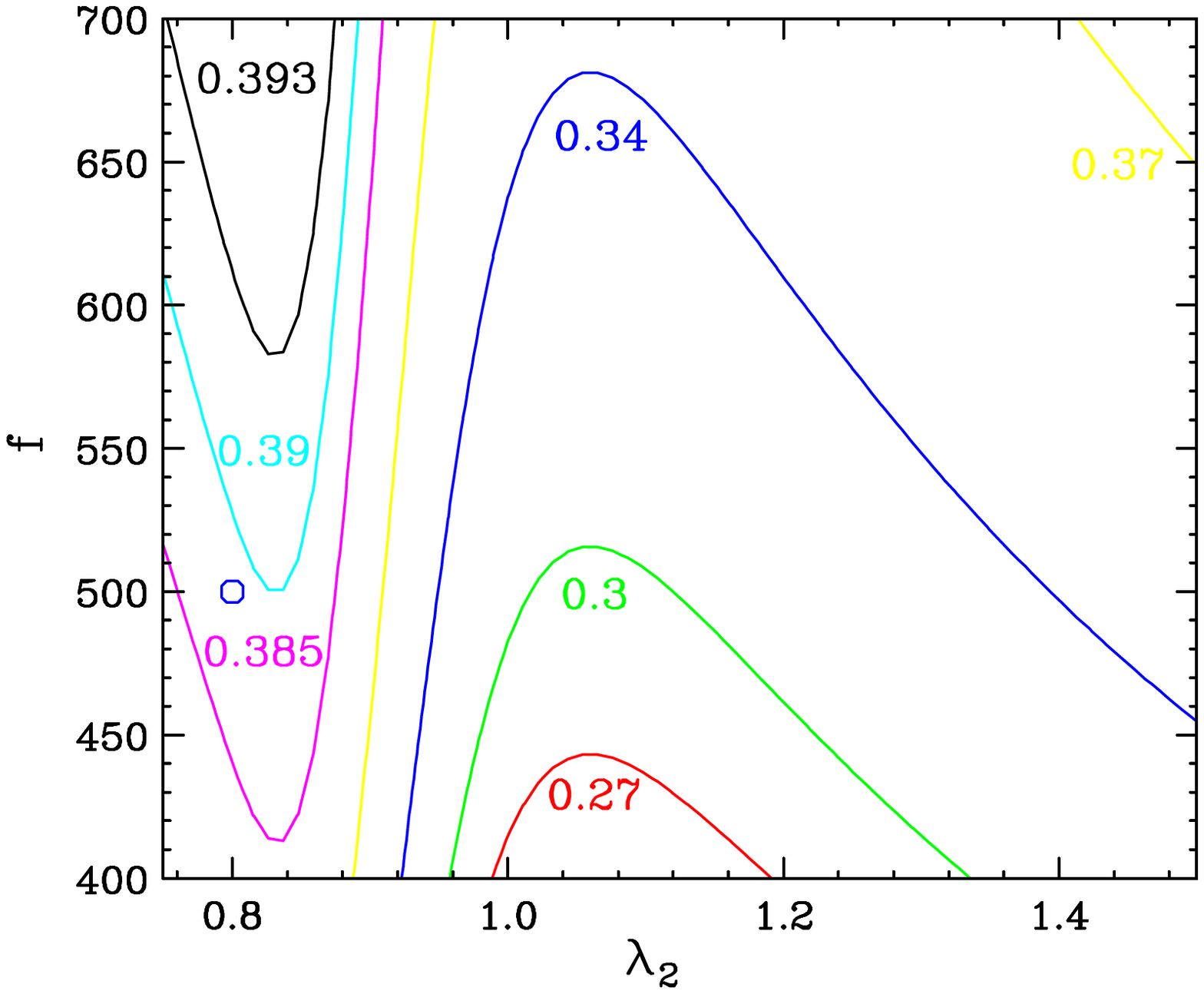} }
  \caption{(a) $A_b$ as a function of $\tan\theta_{L/R} = f_L/f_R$ in SUSY (blue) and LHT (red).
The masses of top-partner ($\tilde{t}$ or $T$) and LSP ($\tilde{\chi}_1^0$)/LTP ($A_H$)
are taken to be
$M_T / M_{\tilde{t}} =300 $ GeV and $M_{\tilde{\chi}_1^0}/M_{A_H}= 100$ GeV (solid line),
$M_T / M_{\tilde{t}} =350 $ GeV and $M_{\tilde{\chi}_1^0}/M_{A_H}= 100$ GeV (dash-dotted line),
$M_T / M_{\tilde{t}} =400 $ GeV and $M_{\tilde{\chi}_1^0}/M_{A_H}= 200$ GeV (dashed line), and
$M_T / M_{\tilde{t}} =400 $ GeV and $M_{\tilde{\chi}_1^0}/M_{A_H}= 66.76$ GeV (dotted line).
(b) $A_b$ in LHT in the plane of two independent parameter, $f$ and $\lambda_2$.
The dot represent our study point.}
  \label{angle_b}
\end{figure}
For the process $T\to t A_H \to W^+ b A_H$, the angular distribution of the b-jet in the rest frame
of the top quark is given by
\begin{equation}
\frac{1}{\Gamma_T} \frac{d \Gamma_T}{d\cos\theta_{tb}} \equiv
\frac{1}{2} \big ( 1 + A_b \cos\theta_{tb}\big ) \, ,
\end{equation}
where $\theta_{tb}$ is the angle between spin polarization of top and bottom quark
in the rest frame of top quark, and the coefficient of $A_b$ is related to the spin
vector $S^\mu$ via
\begin{equation}
A_b = - \frac{m_t^2 - 2 m_W^2}{m_t^2 + 2 m_W^2} \hat{S}_t \, ,
\end{equation}
with
\begin{eqnarray}
\hat{S}_t &=& \frac{2m_t \left ( M_T^2 - m_t^2 + 2M_{A_H}^2\right )}{M_{A_H^2} N} |\vec{P}_{A_H}|
\left ( |f_L|^2 - |f_R|^2 \right ) \nonumber \\
&=&\frac{ \left ( |f_L|^2 - |f_R|^2 \right ) }{M_{A_H^2} N}
\left ( M_T^2 - m_t^2 + 2M_{A_H}^2\right ) \nonumber \\
&& \hspace{2cm}\times
\sqrt{ \Big (M_T^2 - (m_t - M_{A_H})^2 \Big ) \Big ( M_T^2 - (m_t + M_{A_H})^2\Big )} \, .
\end{eqnarray}
Here $\vec{P}_{A_H}$ is the three momentum of the LTP in the rest frame of top quark and
$S_t^2 = - S^\mu S_\mu$. In the same way as SUSY, $A_b$ depends on $f_L/f_R$.
In Fig.~\ref{angle_b}, we show $A_b$ with several values of masses and compare with results in SUSY.
For SUSY, $A_b$'s are calculated with formulas given in~\cite{Kitano:2002ss}.
We notice that $A_b$ in LHT is as sensitive
as $A_b$ in SUSY, which has strong dependence on the ratio $f_L/f_R$,
although they have different spin structure in the vertex, $T\to t A_H$.
The statistical uncertainties in the measurement of $A_b$ is even better in the LHT since
the production cross section is much larger than one in SUSY for given masses
(see~\cite{Kitano:2002ss} for detail.).

\section{Summary and conclusions}

In this paper, we studied the phenomenology of top partners in the
context of the MSSM and LHT. In the MSSM, a discrete symmetry,
R-parity, is introduced and the lightest super-particle is
automatically stable. Its characteristic missing energy signals
has been regarded as a genuine feature of the theory. Exactly the same signature
can happen in the LHT where a discrete symmetry, dubbed
T-parity, can play similar roles with those of the R-parity in the
MSSM. Since the collider signals of both theories can mimic each
other's, one need to compare their phenomenology precisely to
understand the underlying physics.

We first estimated the total cross section of the pair production
of top partners (the lightest scalar top quark in the MSSM and the
T-odd top quark partner in the LHT) in the near future linear
colliders where we can have better resolution for various
observable quantities than that of the LHC and better chance to
discriminate the ``{\it fake}'' theories. The larger cross section
is expected for T-odd quark production than scalar top quark
production once their masses are set to be exactly the same since
T-odd quark has 2 times larger number of helicity degrees of
freedom than that of scalar top quark (see Fig.4). Also the
angular distributions for T-odd quark (or its leptonic decay
products) can be distinguishable from that of scalar top quark,
again thanks to its fermionic characteristics (see Fig.5.). The
standard model background was also estimated and we found that it
is quite plausible for us to be able to distinguish the signals of
LHT and MSSM from the ones of the standard model. We provide the
analytic expressions for reconstruction of Top-partner's mass in
the LHT and MSSM in the hadronic or semi-leptonic channel. The
energy distribution of top quark (see Fig. 6) can be a useful
observable quantity for study of top partners. The angular
distribution of b-jet in the final state was studied since it can
provide an important understanding about the helicity of the
top-partner (see Fig.7). In the appendix, we presented asymmetries
for the LHT ($A_{LR}$ and $A_{FB}$). They are found to be useful to
understand the `vectorlikeness' of the T-odd quark in the LHT
using asymmetries.

In conclusion, linear collider experiments with the better
resolution than that of hadronic collider experiments can provide
a nice chance to probe and discriminate the competing theories
beyond the standard model such as the LHT and MSSM which share
similar phenomenological features.

\section{Appendix}
\label{appendix}
\subsection{Left-right polarization asymmetry: $A_{LR}$}
\label{lrsymmetry}
With polarized electron beams the cross sections $\sigma_L$ and
$\sigma_R$ for the scattering of left-handed and right-handed
electrons on unpolarized positrons can be separately measured. The
left-right polarization asymmetry $A_{LR}$ is defined as
\begin{eqnarray}
A_{LR}^{LHT} &=& \frac{\sigma_L - \sigma_R}{\sigma_L+\sigma_R} \\
 &=& \frac{\big ( |G_{LL}|^2+|G_{LR}|^2-|G_{RR}|^2-|G_{RL}|^2 \big )
           \big ( 1+\frac{1}{3}\beta^2\big )
              + 2 Re\big ( G_{LL}G^*_{LR}-G_{RR}G^*_{RL} \big )
             \big ( 1-\beta^2\big ) }
            { \big ( |G_{LL}|^2+|G_{LR}|^2+|G_{RR}|^2+|G_{RL}|^2 \big )
              \big ( 1+\frac{1}{3}\beta^2\big )
              + 2 Re\big ( G_{LL}G^*_{LR}+G_{RR}G^*_{RL} \big )
              \big ( 1-\beta^2\big )  } \nonumber \\
&=& \frac{ \big( {\cal A-B}\big ) \big ( 1+\frac{1}3{}\beta^2\big
)+   {\cal D} \big ( 1-\beta^2\big ) } {\big( {\cal A+B}\big )
\big ( 1+\frac{1}3{}\beta^2\big )+   {\cal C} \big ( 1-\beta^2\big
)} \, ,
\end{eqnarray}
where ${\cal D} = 2 Re\big ( G_{LL}G^*_{LR}-G_{RR}G^*_{RL} \big
)$.
\begin{figure}[t]
\centerline{
  \includegraphics[width=.49\linewidth]{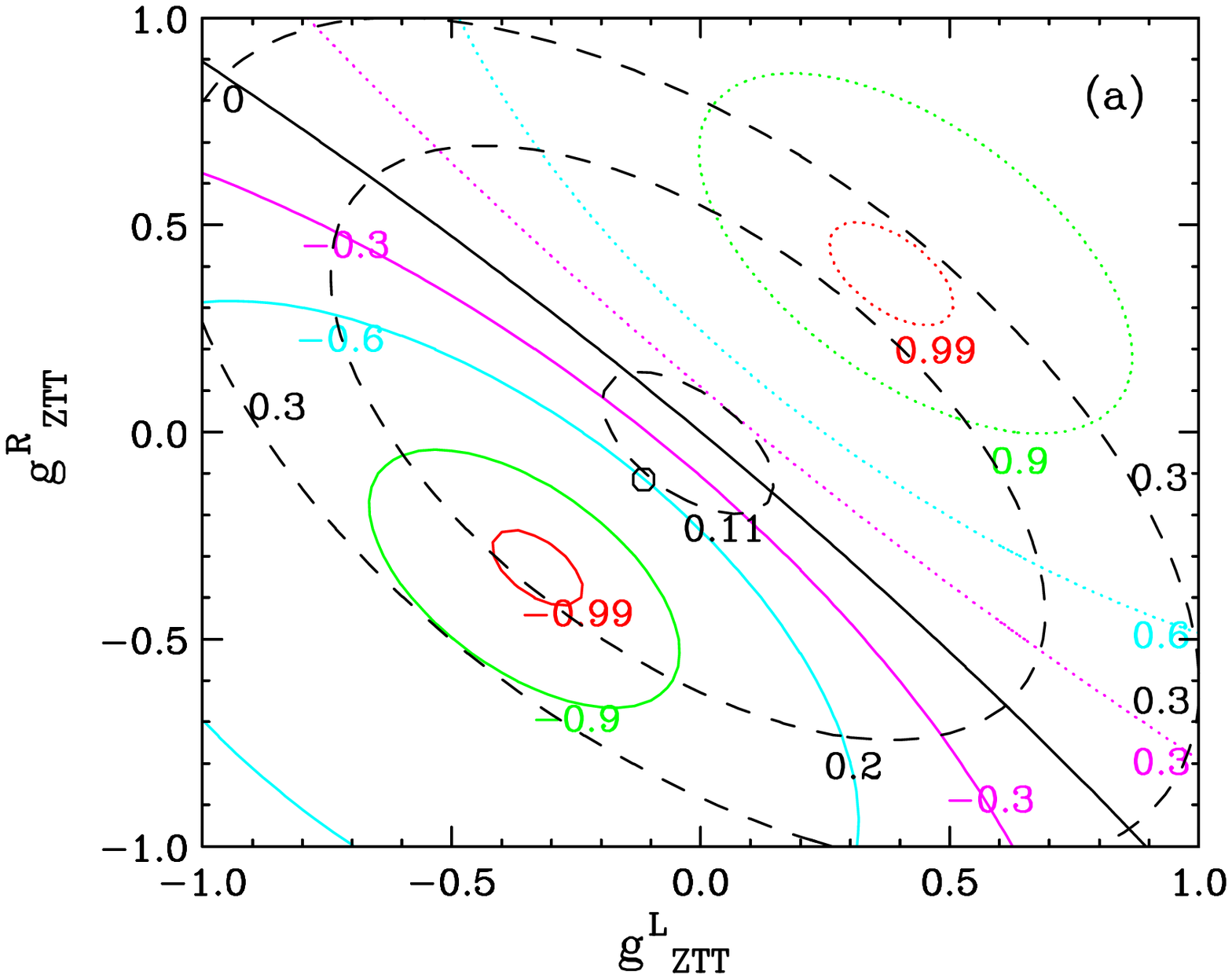}
  \includegraphics[width=.49\linewidth]{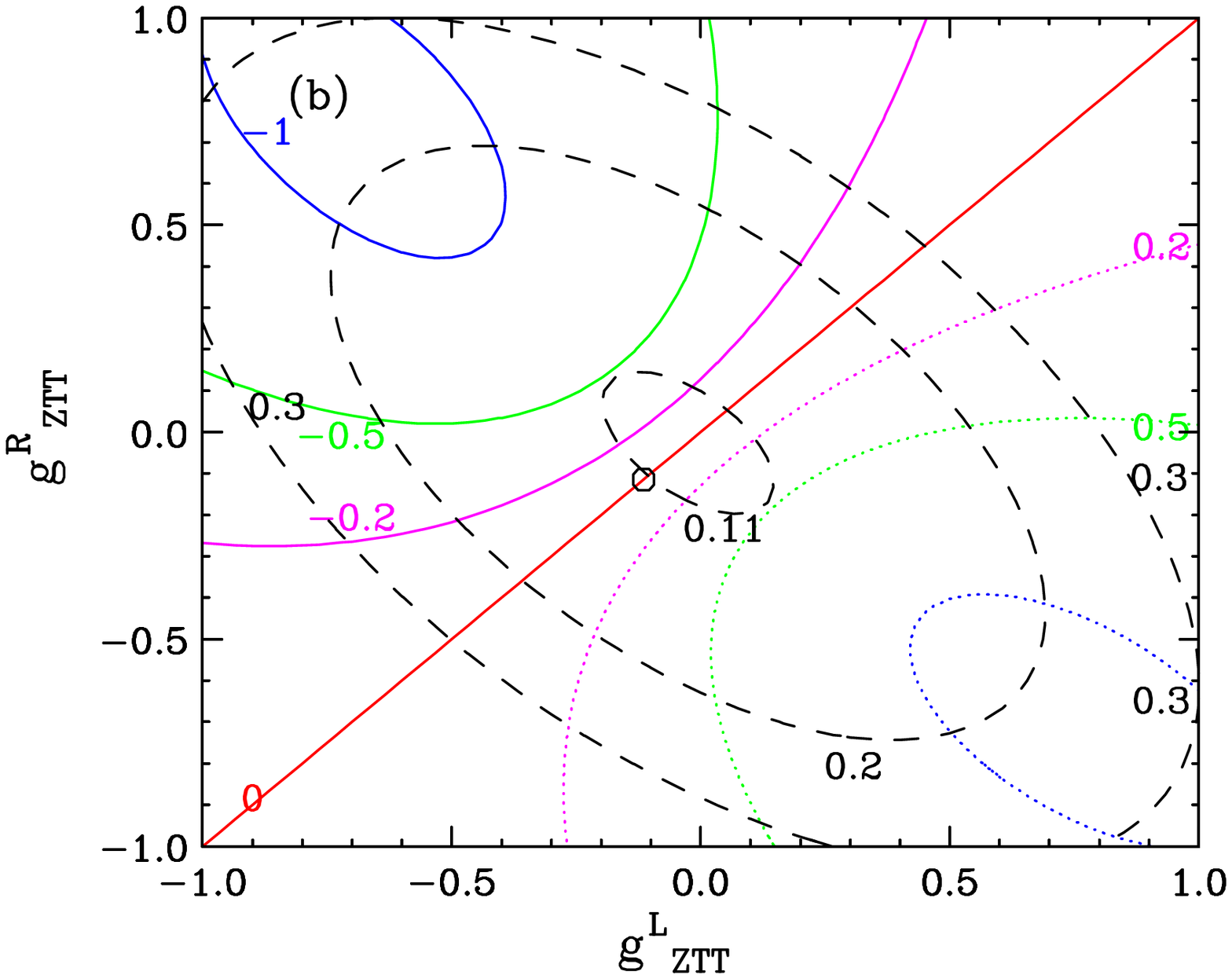} }
  \caption{(a) The left-right asymmetry for polarized scattering and (b)
the forward-backward asymmetry. Here we think of more generic case
  where the vector-like condition or ($G_{LL}=G_{LR}$ and $G_{RR}=G_{RL}$)
does not apply. For LHT, we get $A_{LR}= -0.6$ and $A_{FB}=0$.}
  \label{a_lr}
\end{figure}
Fig.~\ref{a_lr}(a) shows contour lines for several values of
$A_{LR}$ and the dot represent the prediction of LHT,
$g^L_{ZT\bar{T}} = g^R_{ZT\bar{T}} = -\frac{2}{3}
\frac{g}{c_W}s_W^2$. In this plot, we assume that the photon
coupling is purely vector-like but we do not assume $G_{LL} =
G_{LR}$ and $G_{RR} = G_{RL}$. Instead we take the coupling
$\bar{T}_- \gamma_\mu ( g^L_{ZT\bar{T}}P_L + g^R_{ZT\bar{T}}P_R)
Z^\mu T_-$ and vary $g^L_{ZT\bar{T}}$ and $g^R_{ZT\bar{T}}$. Solid
lines represent negative asymmetries: $A_{LR}=-0.99, -0.9, -0.6$
and $-0.3$ while black solid line shows zero asymmetry and dotted
lines represent positive asymmetries: $A_{LR}=0.99, 0.9, 0.6$ and
$0.3$. The total cross section is independent measurement and can
be used to identify the couplings with the aid of this asymmetry.
Dashed contour lines represent cross sections in pb.
\subsection{Forward-backward Asymmetry: $A_{FB}$}
\label{fbsymmetry}
The forward-backward asymmetry for unpolarized beams is defined to
be the number of $T$ at CM scattering angle $\theta$ minus the
number of $T$ at angle $\pi-\theta$ divided by the sum,
\begin{equation}
A_{FB}(\theta) = \frac{d\sigma(\theta) -
d\sigma(\pi-\theta)}{d\sigma(\theta) + d\sigma(\pi-\theta)} \, .
\end{equation}
The integrated asymmetry is
\begin{eqnarray}
A_{FB}^{LHT} &=& \frac{ \int_{0}^{\pi/2}
          \big [ d\sigma(\theta) - d\sigma(\pi-\theta) \big ]}{\sigma} \\
&=& \frac{N({\rm forward}) - N({\rm backward})}{N({\rm forward}) + N({\rm backward})}\\
&=& \frac{ \big ( |G_{LL}|^2+|G_{RR}|^2-|G_{LR}|^2-|G_{RL}|^2 \big
) \beta } { \big ( |G_{LL}|^2+|G_{LR}|^2+|G_{RR}|^2+|G_{RL}|^2
\big ) \big ( 1+\frac{1}{3}\beta^2\big )
              + 2 Re\big ( G_{LL}G^*_{LR}+G_{RR}G^*_{RL} \big )
                    \big ( 1-\beta^2\big )  } \nonumber \\
&=& \frac{{\cal A} - {\cal B}} { \big ( {\cal A} + {\cal B} \big )
(1+\frac{1}{3}\beta^2) + {\cal C} \big ( 1-\beta^2\big )} \, .
\end{eqnarray}
$A_{FB}=0$ (see Fig.~\ref{a_lr}(b)) since $\gamma$ and $Z$
couplings to $T\bar{T}$ are purely vector-like and hence ${\cal A
= B}$.

\subsection{$T_+$ production at the ILC}
\label{teven}

In this appendix, we consider possible production and detection of even parity top partner
for the completeness. First, even parity top partner ($T_+$) is always heavier than
odd parity one ($T_-$).
\begin{eqnarray}
\frac{M_{T_-}}{M_{T_+}}=\frac{\lambda_2}{\sqrt{\lambda_1^2+\lambda_2^2}} \leq 1.
\end{eqnarray}
Contrast to the case at the LHC, where $T_+$ could be singly produced via t-channel W boson exchange
($q b \rightarrow q' T_+$) \cite{Han:2003wu, Perelstein:2003wd}, $T_+$ can only be produced by pair via s-channel photon and
Z boson exchanges at the ILC. Here is the relevant Feynman rules describing $\gamma \bar{T}_+ T_+$ and $Z \bar{T}_+ T_+$ couplings.
\begin{eqnarray}
{\cal L}= \bar{T}_+ \left[\frac{2}{3}\gamma_\mu\left( e A^\mu +(-\frac{ g}{c_w}s_w^2+\frac{3}{4}\frac{g}{c_w}\frac{v^2}{f^2}c_\lambda^4 P_L)Z^\mu\right)\right]T_+,
\end{eqnarray}
where $c_\lambda = \lambda_1/\sqrt{\lambda_1^2+\lambda_2^2}$ was introduced earlier.
$T_+$ mostly decays to $W b$ but it can also decay to $ t H$ and $ T_- A_H$ \cite{Hubisz:2004ft}.

The cross section for $\bar{T}_+T_+$ production is kinematically suppressed in comparison with that of $\bar{T_-}T_-$
\begin{eqnarray}
\frac{\sigma_{e^+e-\rightarrow \bar{T}_+ T_+}}{\sigma_{e^+e^-\rightarrow \bar{T}_- T_-}}
 \simeq \frac{\sqrt{1-4 M_{T_+}^2/{s}}(1+2 M_{T_+}/{s})}{\sqrt{1-4 M_{T_-}^2/{s}}(1+2 M_{T_-}/{s})},
\end{eqnarray}
with additional $\frac{v^2}{f^2}c_\lambda^4$ order corrections. For the linear collider with
$\sqrt{s}= 1$ TeV, it is challenging to see the signals of T-even top quarks since they only
can be produced near the threshold. Only very restrictive parameter space is
available for $M_{T_+} \leq 500$ GeV.

\begin{acknowledgments}
We thank A. Pukhov and F. Maltoni for useful
correspondence regarding many questions about {\tt CalcHEP} and
{\tt MadGraph/MadEvent}. We are also grateful to C.P. Yuan and
R. Kitano for discussion about the spin vector method,
and M. Perelstein for reading our manuscript and useful comments.
Fermilab is operated by Fermi Research Alliance, LLC under
Contract No. DE-AC02-07CH11359 with the United States Department
of Energy. SP is supported by BK21 program of Korean government.
\end{acknowledgments}

\end{document}